\documentclass[aps,prb,superscriptaddress,twocolumn]{revtex4-1}
\pdfoutput=1
\usepackage{bm}
\usepackage{amstext}
\usepackage{amsmath}
\usepackage{graphicx}
\usepackage{enumerate}
\usepackage{color}
\usepackage{hyperref}
\hypersetup{colorlinks=true,linkcolor=red,citecolor=blue,urlcolor=blue}

\newcommand{\pd}{\partial}

\begin{document}

\title{Reply to 'Comment on "Thermodynamics of quantum crystalline membranes"'}

\author{B. Amorim}
\email[Electronic address: ]{amorim.bac@icmm.csic.es}
\affiliation{Instituto de Ciencia de Materiales de Madrid, CSIC, Cantoblanco
E28049 Madrid, Spain}

\author{R. Rold\'an}
\email[Electronic address: ]{rroldan@icmm.csic.es}
\affiliation{Instituto de Ciencia de Materiales de Madrid, CSIC, Cantoblanco
E28049 Madrid, Spain}

\author{E. Cappelluti}
\affiliation{Istituto dei Sistemi Complessi, CNR, U.O.S. Sapienza, v. dei
Taurini 19, 00185 Roma, Italy}

\author{F. Guinea}
\affiliation{Instituto de Ciencia de Materiales de Madrid, CSIC, Cantoblanco
E28049 Madrid, Spain}

\author{A. Fasolino}
\affiliation{Radboud University Nijmegen,Institute for Molecules and Materials,
NL-6525AJ Nijmegen, The Netherlands}

\author{M. I. Katsnelson}
\affiliation{Radboud University Nijmegen,Institute for Molecules and Materials,
NL-6525AJ Nijmegen, The Netherlands}

\date{\today}

\begin{abstract}
In this note, we reply to the comment made by E.I.Kats and  V.V.Lebedev [arXiv:1407.4298] on our recent work "Thermodynamics of quantum crystalline membranes" [Phys. Rev. B \textbf{89}, 224307 (2014)]. Kats and Lebedev question the validity of the calculation presented in our work, in particular on the use of a Debye momentum as a ultra-violet regulator for the theory. We address and counter argue the criticisms made by Kats and Lebedev to our work.
\end{abstract}

\maketitle

\section{Introduction: Summary of our work and main criticisms presented by Kats and Lebedev}

We begin by briefly summarizing our recent work, Ref.~\onlinecite{Amorim_2014}.
Our aim was to study the thermodynamic properties of crystalline
membranes in the low temperature limit, where quantum effects dominate.
In particular, our main goals were to determine the low temperature
behavior of the thermal expansion and specific heat of a crystalline
membrane, and to estimate the crossover temperature above which quantum
effects are negligible and the classical theory can be safely used.
In order to achieve that, our starting point was the standard, classical,
anharmonic, continuous theory of crystalline membranes of Nelson \&
Peliti\cite{NP87} (which is based on the same Hamiltonian as the usual plate theory\cite{LL59}). 

\begin{figure}
\begin{centering}
\includegraphics[width=8cm]{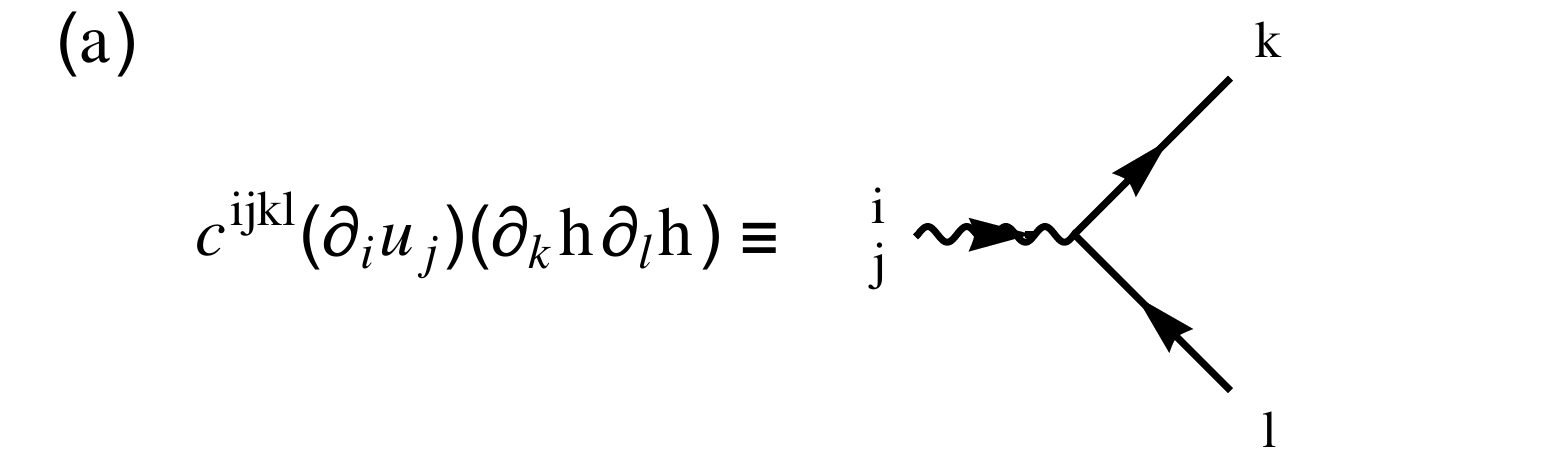}

\includegraphics[width=8cm]{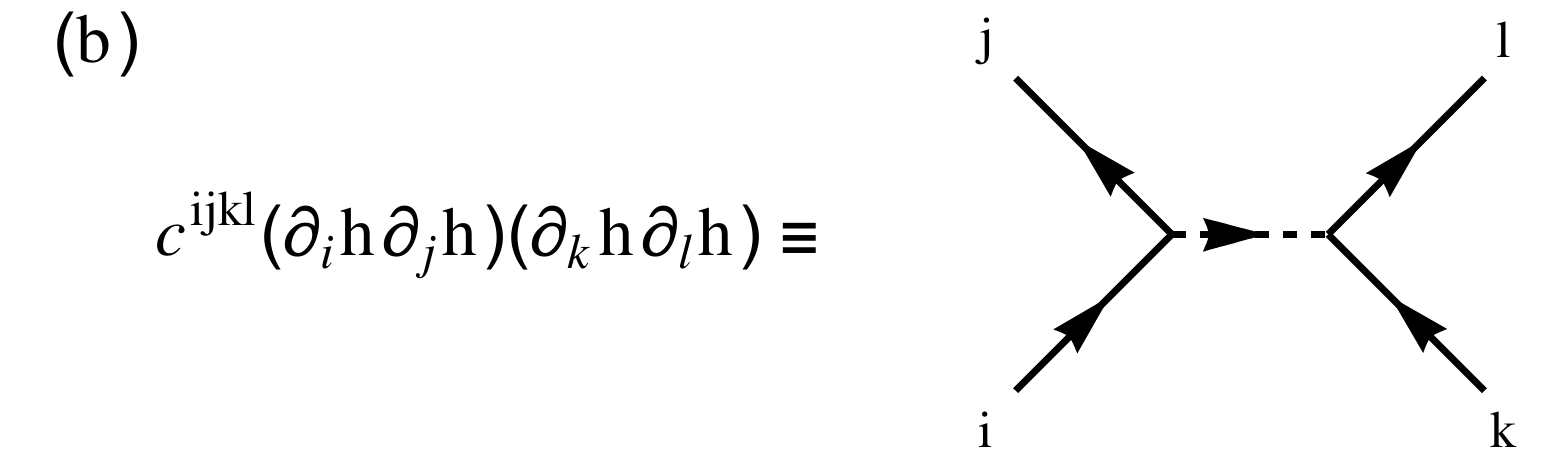}
\par\end{centering}

\caption{\label{fig:Diagram_int}Diagrammatic representation of the interaction
terms in the standard theory for crystalline membranes. Straight lines
represent the out-of-plane propagator, wiggly lines represent the in-plane
propagator. The dashed line represents the quartic interaction for
the out-of-plane mode.}
\end{figure}

In the standard theory of membranes, the deviations of the point mass
positions from the flat configuration are described in terms of an
in-plane displacement, $\vec{u}$, and an out-of-plane displacement,
$h$. In our theory (and in the standard classical theory \cite{NP87}),
there is a cubic interaction between the in-plane and out-of-plane
displacements, of the generic form $\pd u(\pd h)^{2}$, and a quartic
interaction involving only the out-of-plane displacement, of the form
$(\pd h)^{4}$, see Figure~\ref{fig:Diagram_int}.

We have quantized this classical theory by using the Feynman
path integral formalism in imaginary time. By integrating out the
in-plane displacements, we obtain an effective action for the out-of-plane
displacement. By computing, to first order in perturbation theory,
the self-energy for the out-of-plane displacement we obtained, in
the long-wavelength limit, a contribution that goes as $k^{2}$ ($k$
is the momentum). It is important to notice that the bare action
does not contain such term, but only a $k^{4}$ term (associated with
a bending energy). We emphasize that the $k^{2}$ contribution to
the self-energy is only obtained if effects of retardation of the
in-plane phonon in the interactions $\pd u(\pd h)^{2}$ are taken
into account. 

In order to perform the calculation, we have regularized the theory
by adding a high momentum cutoff, above which the continuum theory
we are employing breaks down. We made a "natural" identification
of this cutoff as the Debye momentum, $q_{D}\sim1/a$ where $a$ is
the lattice spacing of the crystalline membrane.

The criticism expressed by Kats and Lebedev in Ref.~\onlinecite{KL_14c} (which had already been made in  Ref.~\onlinecite{Kats_2013}) goes along two main lines: 

\begin{enumerate}[(a)]
\item such a $k^{2}$ contribution is a tension term and therefore should
be zero for a free membrane;
\item it is the use of the "natural" cutoff that leads us to this
(in the point of view of Kats and Lebedev) wrong term. 
\end{enumerate}

The remaining of this note will be organized as follows. In Section~\ref{sec:criticisms}, we will answer to the criticism made by Kats and Lebedev in Ref.~\onlinecite{KL_14c}. We will argue that the $k^{2}$ contribution is
not in reality a tension, but instead a renormalization of the bending
rigidity of the membrane, that acquires a non-trivial momentum dependence. This is essentially the same situation as in the classical theory \cite{NP87}.
We will also argue that the use of the "natural" cutoff, although
only an approximation, allows us to take into account the contribution
of modes in the whole Brillouin zone of the crystalline membrane at
a level that is sufficient for the tasks we set ourselves to accomplish
in our work. In Section~\ref{sec:comparison}, we will briefly discuss and compare the work by Kats and Lebedev Phys. Rev. B \textbf{89}, 125433 (2014), Ref.~\onlinecite{Kats_2013}, also on the problem of quantum crystalline membranes, with ours. It is not the purpose of this note to make a criticism or a detailed analysis of Ref.~\onlinecite{Kats_2013}. However, since the results presented there seem to be at odds with the results presented by us in Ref.~\onlinecite{Amorim_2014}, we believe the reader will benefit from such discussion.

\section{Answer to the criticisms}\label{sec:criticisms}

\subsection{On the use of a "natural" cutoff}

Let us first analyze the last point (b). Kats and Lebedev argue that
from a renormalization group point of view, any high momentum divergence
that appears in the calculation of some physical quantity, should
be absorbed into a redefinition of the bare parameters of the action.
The exact prescription in which these redefinitions should be performed
are usually referred to as renormalization conditions. For this particular
case, Kats and Lebedev argue the renormalization condition is that
the membrane should be under zero tension, that is the $k^{2}$ term that we
obtained should be canceled by adding a similar extra term in the
bare action. 

First of all, we would like to comment that this general regularization
and renormalization procedure can only be performed for a limited
number of theories, which are referred to as renormalizable. Renormalizability
is a very desirable property of a field theory, since for such theories,
once a finite number of parameters are determined by experiments,
all other quantities can be unambiguously computed. In this sense,
renormalizable theories have predictive power. It is generally postulated,
that fundamental theories of nature should be renormalizable.

In condensed matter, continuum field theories usually arise as long-wavelength
approximations to a more fundamental and complete theory that is generally
known. Since the more complete theory is known, we also know the range
of validity of the field theory. In general, the field theory will
only be valid for momenta smaller than $\sim1/a$, where $a$ is a
lattice spacing. If one is lucky, the obtained continuum field theory
is renormalizable, and one can use the well known machinery of renormalization
to study it. However, nothing ensures that the continuum field theory
will be renormalizable, and very often, it is not.

Even if it turns out that the long-wavelength field theory is non-renormalizable,
that does not mean that it is useless. Long-wavelength, continuous
field theories can still be useful in order to study effects which
would be computationally intractable if one were to use a more complete
theory (such as an atomistic model or an \textit{ab initio} method).
In this situation, the parameters to be used in the bare action of
the field theory are to be fed from the calculations using the more
complete theory (in a certain approximation which does not capture
the effect we are interested in). Then we use the field theory in
order to study such effects. While using the field theory, one will
be generally faced with divergent contributions due to high momentum.
This just means that modes with all momenta (modes over all the Brillouin
zone) will contribute to a given quantity. Although the field theory
is, strictly speaking, not valid at high momenta, the contribution
from high momentum modes can be estimated using a high momentum cutoff
of the order of $\sim1/a$, the Debye momentum. 

This last approach, is the approach we employ in our work Ref.~\onlinecite{Amorim_2014}.
For the bare parameters of our model, we use values obtained using
an atomistic, classical model \cite{FLK07,ZKF09}. Then, we use a
continuous field theory in order to study the effects of quantum fluctuations
(which are not taken into account in the classical model) and long
wavelength fluctuations in the thermodynamic limit (the atomistic
model is limited to study finite size systems).

\subsection{On the obtention of a $k^{2}$ contribution}

Now, let us analyze the point (a). First, we would like to point out
that the $k^{2}$ contribution to the out-of-plane mode self-energy
that was found in our calculation is not a tension. Although we state
in our paper
\begin{quote}
"The present result of $\eta=2$ indicates that quantum anharmonic
effects act as an effective positive external strain, which contributes
to the stabilization of the 2D phase of the membrane (see also Ref.
32 ).",
\end{quote}
the view of the $k^{2}$ term as a tension/strain is to be understood only
as an analogy (since a tension always gives origin to a $k^{2}$ contribution).
As a matter of fact, the $k^{2}$ behaviour is somewhat of a coincidence.
In general, and as stated in our paper, we will obtain a $k^{4-\eta}$
behaviour for the self-energy. This just means that the bending rigidity
of the membrane will acquire a dependence on momentum $\kappa(k)\sim k^{-\eta}$, where $\eta$ is some characteristic exponent.
It is only at the level we solved the theory (first order perturbation
theory and self-consistent calculation neglecting corrections to the
in-plane correlators) that the $\eta=2$ is obtained in the quantum
problem at zero temperature. As we say in Ref. \onlinecite{Amorim_2014}, for a more complete calculation we expect
the value of $\eta$ to be changed to a some other value different,
but close, to $2$. 

It is worthwhile comparing our results with the results obtained in
the classical theory for crystalline membranes. We start noting that,
by taking the classical limit (formally setting all Matsubara frequencies
to zero) of the effective action for our quantum theory (equation
13 of Ref.~\onlinecite{Amorim_2014}), one obtains the classical
action from the paper by Nelson \& Peliti, Ref.~\onlinecite{NP87}. If
one starts from this classical action and performs a first order calculation
for the out-of-plane self-energy, one obtains \cite{NP87,NPW04}
\begin{align}
\Sigma_{k} & = \frac{4\mu(\lambda+\mu)}{\lambda+2\mu}k_{B}T\int\frac{d^{2}q}{\left(2\pi\right)^{2}}\frac{\left[\vec{k}\times\vec{q}\right]^{4}}{q^{4}}\frac{1}{\kappa\left|\vec{k}+\vec{q}\right|^{2}} \nonumber \\ 
           & = \frac{4\mu(\lambda+\mu)}{\lambda+2\mu}\frac{3k_{B}T}{16\pi\kappa}k^{2},
\end{align}
(equation 24 of Ref.~\onlinecite{Amorim_2014}). This is a well
known result, which has been used to estimate the momentum scale below
(or, with the replacement $k\rightarrow2\pi/L$, the membrane size
above) which anharmonic effects become dominant (see for instance
equation 11 from Ref.~\onlinecite{NP87}, equation $5.2$ from Ref.~\onlinecite{XML03},
equation 10 from Ref.~\onlinecite{BH10} and equation 40 of Ref.~\onlinecite{Hasselmann_2011}). This is the analog of the Ginzburg criterion for critical phenomena. Notice that this value is obtained for a crystalline membrane in the
absence of any external tension. Nobody in the theory of membranes
has ever claimed, to our knowledge, that this term should be just neglected,
in virtue of the condition of zero surface tension. At $T=0$, we
have the term with the same $k$-dependence. The only difference is
that at $T=0$ it depends on the cut-off. We do not believe that
this difference has any meaning, if we do not postulate that all condensed
matter theories should be renormalizable in a quantum field theory
sense. We are dealing with the theory of anharmonic phonons \cite{C__63}
(for a recent presentation, see Ref.~\onlinecite{KT_02}), and from the very
beginning all summations on the momenta are restricted by the Brillouin
zone. "Inapplicability" of the continuum
medium theory in this situation means that when using the Debye model
for the phonons we are not guaranteed that the numerical factor is
correct (actually, it is not), but this not a reason to say that this factor should be  zero.

Moreover, the condition of zero tension in two dimensions is equivalent
to the condition of zero pressure in three dimensions, and in the
latter case it is well known how to deal with this condition. When
considering thermal expansion in theory of crystals, one needs first
to calculate the phonon contribution to the pressure; nobody has ever
put this correction to zero but use it to calculate the change of
the equilibrium lattice parameter induced by this pressure \cite{KT_02}.
This is exactly how we use this $k^{2}$ term, to find the analog
of the Ginzburg criterion and to calculate the contribution to the
thermal expansion.

\section{Comparison of Phys. Rev. \textbf{B} 89, 125433 (2014) with our work }\label{sec:comparison}

In Phys. Rev. B 89, 125433 (2014) \cite{Kats_2013}, the authors perform
a Wilsonian perturbative renormalization calculation, where fluctuations
are integrated out step-by-step starting from large momentum fluctuations
and going towards small momentum fluctuations. This is to be contrasted
with the approach of our own work Ref.~\onlinecite{Amorim_2014}, where
the in-plane modes are integrated once and for all momenta, and we are left
with an effective theory for the out-of-plane modes. Nevertheless,
we will try to show the correspondence between the two approaches.

Let us start with the standard stretching energy term for the crystalline
membrane
\begin{equation}
\mathcal{U}_{\text{strech}}=\frac{1}{2}\epsilon_{ij}c^{ijkl}\epsilon_{kl},
\end{equation}
where $\epsilon_{ij}=\left( \pd_{i}u_{j} + \pd_{j}u_{i} + \pd_{i}h\pd _{j}h \right )/2$ is the relevant strain tensor and  $c^{ijkl}=\lambda\delta^{ij}\delta^{kl}+\mu\left(\delta^{ik}\delta^{jl}+\delta^{il}\delta^{jk}\right)$
is the elastic moduli tensor for an isotropic membrane. The stretching
energy contains the usual quadratic term for the in-plane displacements,
$\pd_{i}u_{j}c^{ijkl}\pd_{k}u_{l}/2$; a cubic term between in-plane
and out-of-plane displacements, $c^{ijkl}\left(\pd_{i}u_{j}\right)\left(\pd_{l}h\pd_{k}h\right)/2$,
which we represent diagrammatically as in Figure~\ref{fig:Diagram_int}(a);
and a quartic term for out-of-plane displacements, $c^{ijkl}\left(\pd_{i}h\pd_{j}h\right)\left(\pd_{l}h\pd_{k}h\right)/8$,
which we represent diagrammatically in Figure~\ref{fig:Diagram_int}(b). 

\begin{figure}
\begin{centering}
\includegraphics[height=4cm]{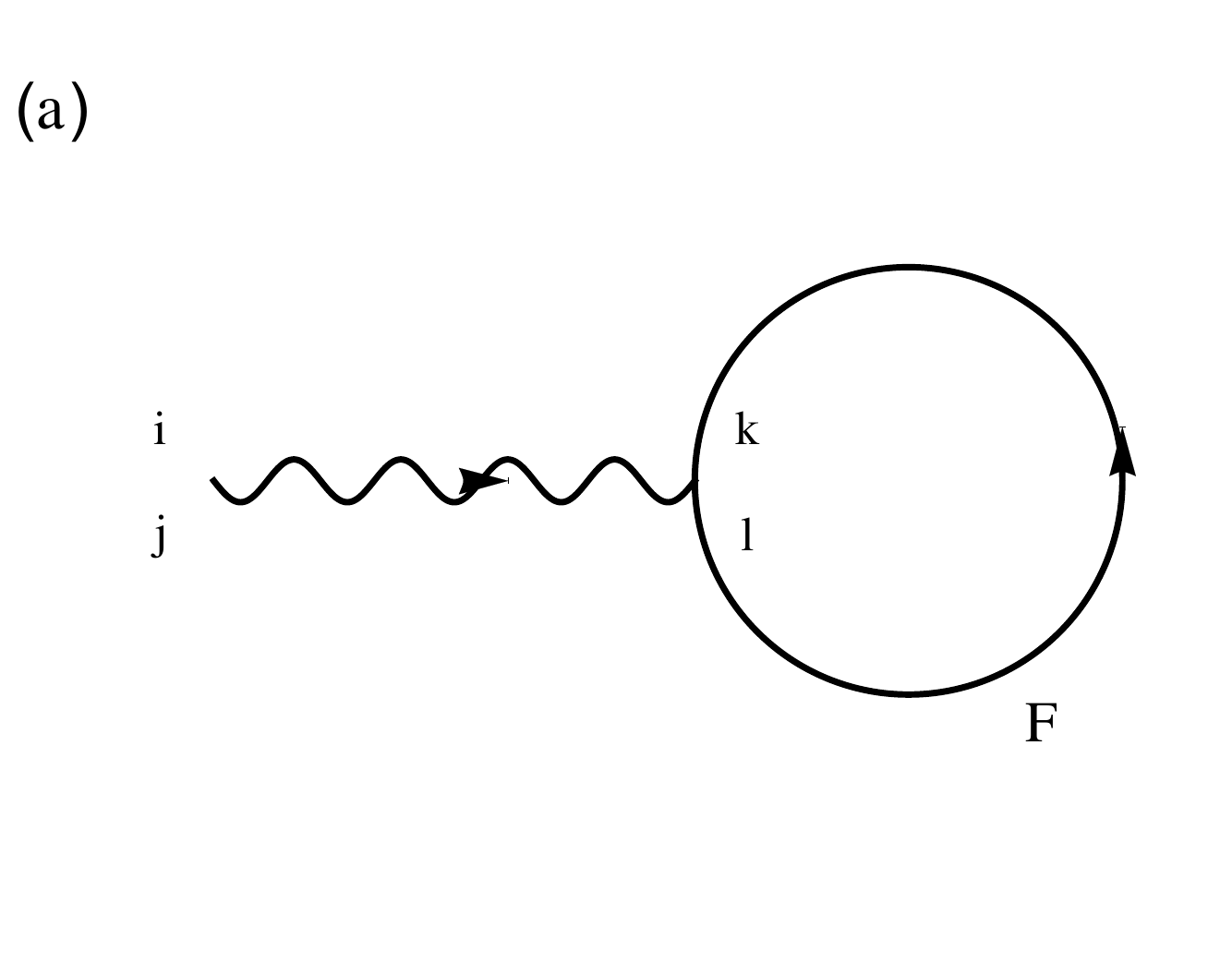}
\par\end{centering}

\caption{\label{fig:u}Diagram that contributes to the generation of a linear
term in $\pd_{i}u_{j}$. The $F$ subscript means that the loop involves
fast modes.}
\end{figure}

In a Wilsonian renormalization approach, fields are split between
slow fields (with momenta from $0$ up to $\Lambda^{\prime}$), which
we will denote by $u_{S}$ and $h_{S}$, and fast fields (with momenta
between $\Lambda^{\prime}$and $\Lambda$), which we will denote as
$u_{F}$ and $h_{F}$. When writing the in-plane strain in Fourier
modes one has to treat the homogeneous component separately, such
that we have
\begin{equation}
\pd_{i}u_{j}\left(\vec{x},\tau\right)=u_{ij}^{0}+\frac{1}{\sqrt{\beta V}}\sum_{iq_{n},\vec{q}\neq0}iq_{i}u_{j}\left(\vec{q},iq_{n}\right)e^{i\vec{q}\cdot\vec{x}}e^{-iq_{n}\tau},
\end{equation}
where $u_{ij}^{0}$ is the homogeneous strain component term. Therefore,
$u_{ij}^{0}$ is always a slow variable. After integrating out the
fast variables, the partition function can be written as 
\begin{align}
Z & =\int D\left[u_{S},h_{S},u_{F},h_{F}\right]e^{-S_{\Lambda}\left[u_{S},h_{S},u_{F},h_{F}\right]}\nonumber \\
  & =\int D\left[u_{S},h_{S}\right]e^{-S_{\Lambda^{\prime}}\left[u_{S},h_{S}\right]},
\end{align}
where
\begin{equation}
e^{-S_{\Lambda^{\prime}}\left[u_{S},h_{S}\right]}=\int D\left[u_{F},h_{F}\right]e^{-S_{\Lambda}\left[u_{S},h_{S},u_{F},h_{F}\right]}.
\end{equation}
 When integrating out the fast modes, the interaction pictured in
Figure~\ref{fig:Diagram_int}(a) will generate, at one loop, a linear
term for $u_{ij}^{0}$ of the form of Figure~\ref{fig:u}, which can
be written as 
\begin{equation}
\Delta S_{\Lambda^{\prime}}^{(0)}\left[u_{S}\right]=\tau^{ij}\int_{0}^{\beta}d\tau\int d^{2}x\pd_{i}u_{j},
\end{equation}
with
\begin{equation}
\tau^{ij}=\frac{1}{2}\left(\lambda+\mu\right)\frac{1}{\beta V}\sum_{ip_{n},\vec{p},\Lambda^{\prime}<\left|\vec{p}\right|<\Lambda}p^{2}\left\langle h_{\bm{p}}h_{-\bm{p}}\right\rangle _{0}.
\end{equation}
$\tau^{ij}$ seems to act like an externally applied stress/tension. 

Recall that an in-plane (stretching) stress, $\sigma^{ij}$, couples
to the in-plane strain as a term in the potential energy of the form
$-\pd_{i}u_{j}\sigma^{ij}$. This linear term can be eliminated by
making a shift in the fields, $\pd_{i}u_{j}\rightarrow\pd_{i}u_{j}-c^{ijkl}\sigma_{kl}$,
at the cost of generating a term of the form $\sigma^{ij}\pd_{i}h\pd_{j}h/2$.
Such a term indeed gives origin to a $k^{2}$ term in the self-energy
of the out-of-plane displacement field. 

\begin{figure}
\begin{centering}
\includegraphics[height=4cm]{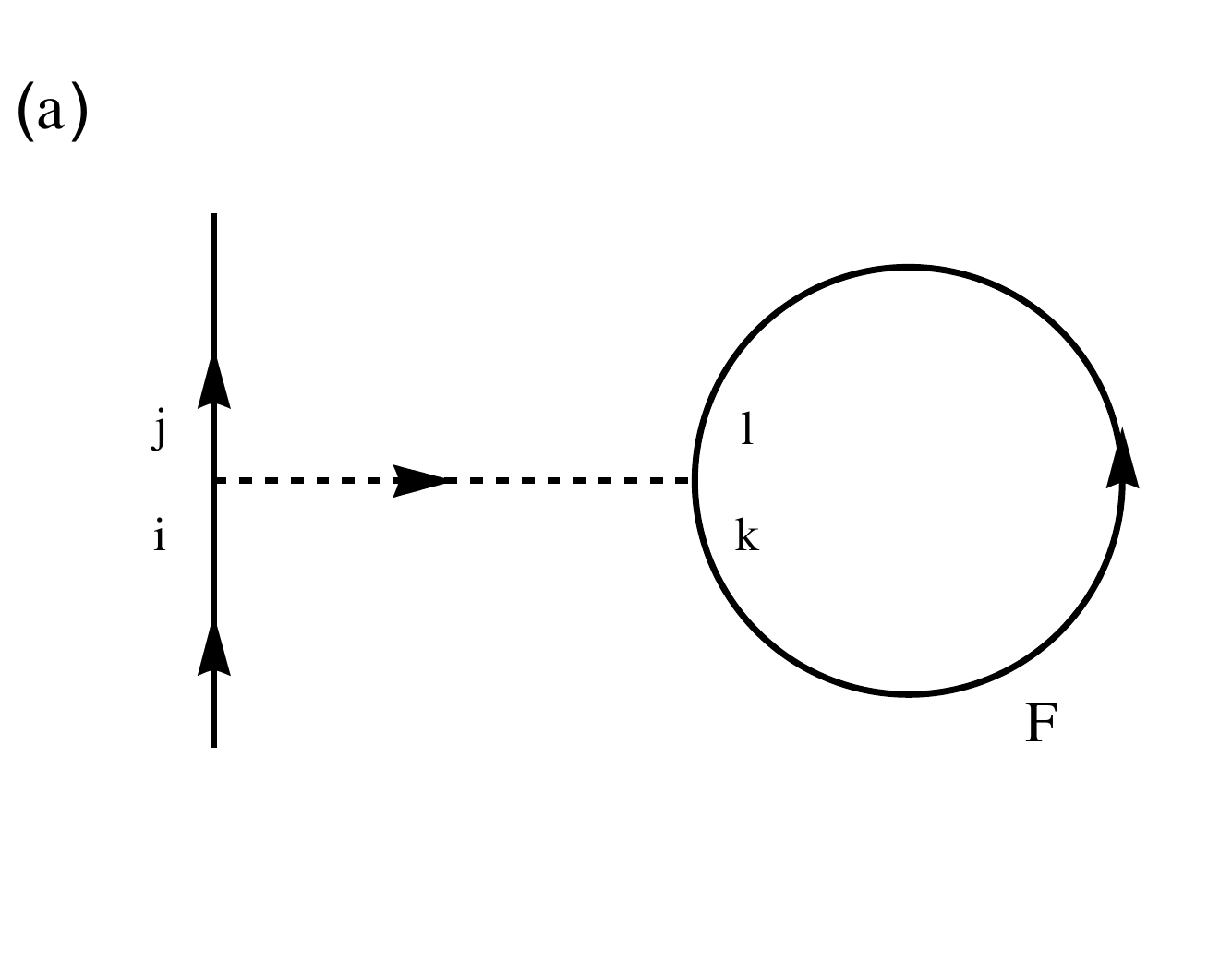}

\includegraphics[height=4cm]{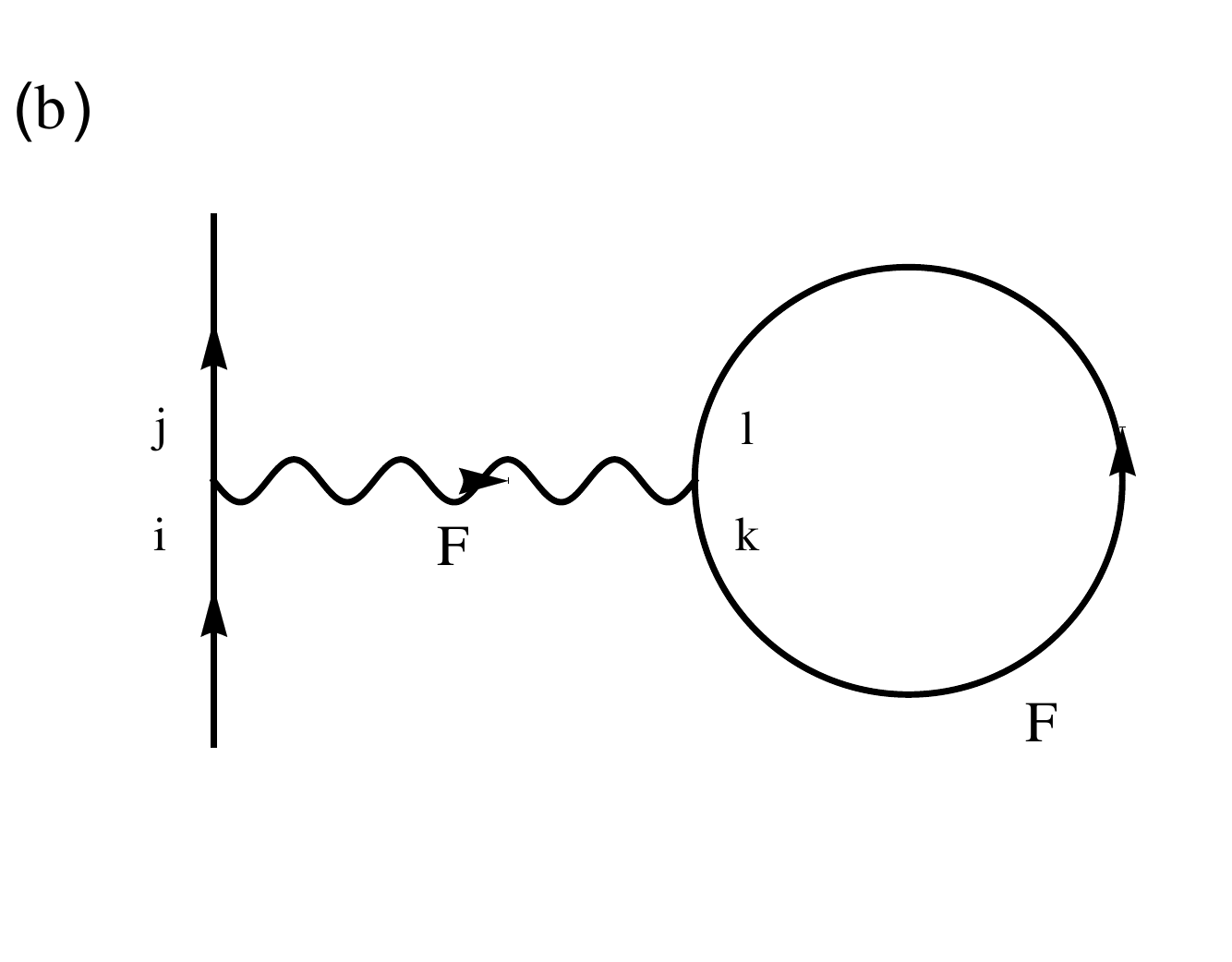}
\par\end{centering}

\caption{\label{fig:Tadpole/Hartree}Tadpole/Hartree diagrams that contribute
to the correction of the out-of-plane quartic term. The diagram (b)
does not contribute when performing a Wilsonian renormalization calculation.}
\end{figure}

However, $\tau^{ij}$ is not a real stress. To see this, one must
notice that besides generating $\Delta S_{\Lambda^{\prime}}^{(0)}\left[u_{S}\right]$,
one will also generate a quadratic term for the out-of-plane mode,
see Figure~\ref{fig:Tadpole/Hartree}(a), which is exactly given
by
\begin{equation}
\Delta S_{\Lambda^{\prime}}^{(1)}\left[h_{S}\right]=\frac{1}{2}\tau^{ij}\int_{0}^{\beta}d\tau\int d^{2}x\pd_{i}h\pd_{j}h.\label{eq:h2_cancelation}
\end{equation}
Therefore, collecting $\Delta S_{\Lambda^{\prime}}^{(0)}\left[u_{S}\right]$
and $\Delta S_{\Lambda^{\prime}}^{(1)}\left[u_{S}\right]$ one obtains
\begin{multline}
\Delta S_{\Lambda^{\prime}}^{(0)}\left[u_{S}\right]+\Delta S_{\Lambda^{\prime}}^{(1)}\left[h_{S}\right] = \\
 = \int_{0}^{\beta}d\tau\int d^{2}x\tau^{ij}\left(\pd_{i}u_{j}+\frac{1}{2}\pd_{i}h\pd_{j}h\right).
\end{multline}
Such a term, indeed does not give origin to a $k^{2}$ term in the
out-of-plane phonon self-energy. To see this, notice that the term
$\tau^{ij}\pd_{i}u_{j}$ can be eliminated by performing a shift in
the fields, $\pd_{i}u_{j}\rightarrow\pd_{i}u_{j}-c^{ijkl}\tau_{kl}$,
at the expense of generating a new term, $-\tau^{ij}\pd_{i}h\pd_{j}h/2$.
This new term, will exactly cancel the term in $\Delta S_{\Lambda^{\prime}}^{(1)}\left[h_{S}\right]$.
Therefore, the dispersion relation of the flexural phonon is left
unchanged by the diagram from Figure~\ref{fig:Tadpole/Hartree}(a).
In the approach employed in our work, Ref.~\onlinecite{Amorim_2014}, this
fact manifests itself by the non-existence of Hartree/tadpole diagrams
in our perturbative calculation (the diagrams from Figure~\ref{fig:Tadpole/Hartree}(a)
and (b) exactly cancel). Furthermore, notice that in a Wilsonian
renormalization calculation, the diagram from Figure~\ref{fig:Tadpole/Hartree}(b) never occurs since, by momentum conservation, the in-plane mode
line necessarily carries zero momentum and therefore is not a fast
variable.

\begin{figure}
\begin{centering}
\includegraphics[height=5cm]{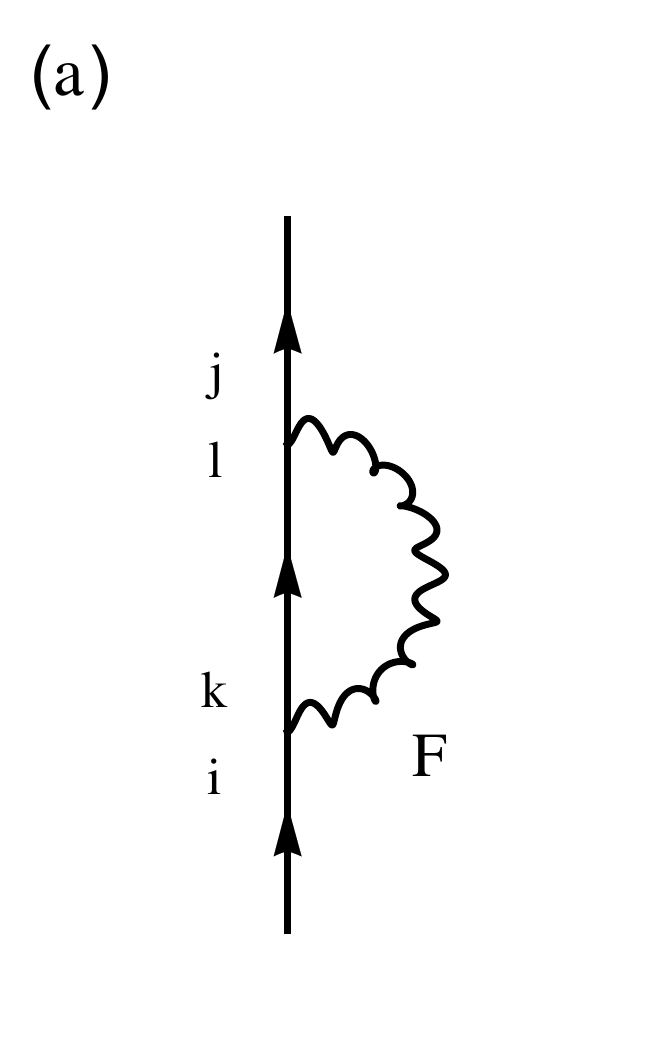}\includegraphics[height=5cm]{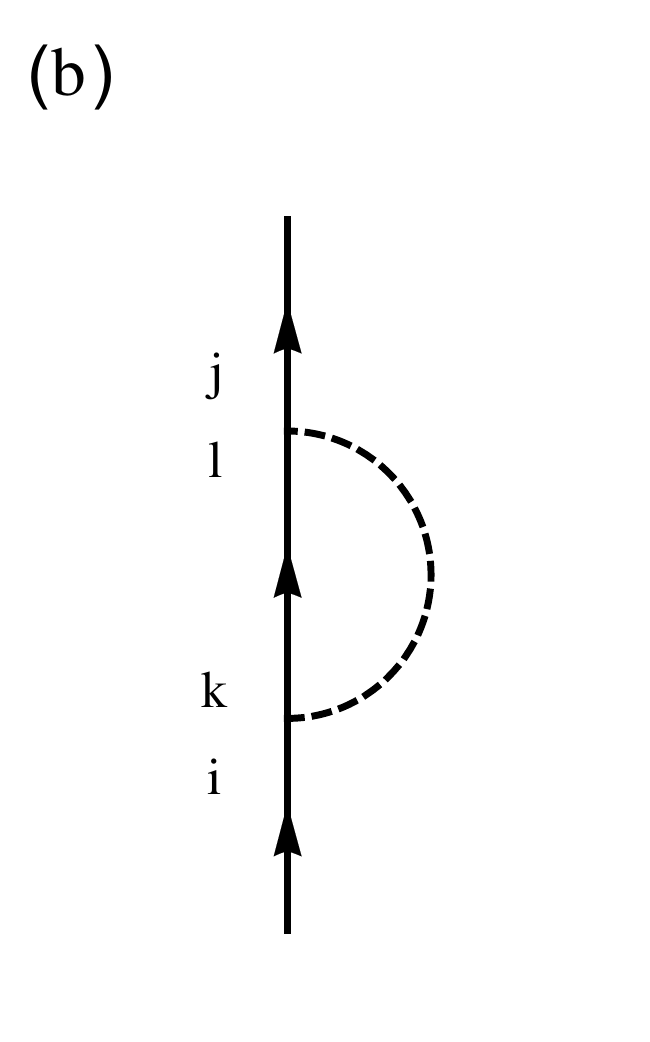}
\par\end{centering}

\caption{\label{fig:Sunset/Fock}Sunset/Fock diagrams that contribute to the
renormalization of the bending rigidity of the membrane.}
\end{figure}

However, besides generating the term $\Delta S_{\Lambda^{\prime}}^{(1)}\left[h_{S}\right]$,
two more terms are generated at one loop, that are quadratic in the
out-of-plane displacement. These two terms are represented by the
diagrams in Figure~\ref{fig:Sunset/Fock}(a) and (b). These are the
diagrams that we consider in our perturbative calculation in Ref.~\onlinecite{Amorim_2014},
and are also the ones that are considered in the classical theory
of crystalline membranes of Nelson \& Peliti \cite{NP87}. Contrary
to the diagram of Figure~\ref{fig:Tadpole/Hartree}(a), these diagrams
do not have a partner diagram giving origin to a linear term in $\pd_{i}u_{j}$,
and therefore, cannot be eliminated with a shift of $\pd_{i}u_{j}$.
Therefore, the diagrams from Figure~\ref{fig:Sunset/Fock}(a) and
(b) will be responsible for a correction to the membrane bending rigidity.

In our work Ref.~\onlinecite{Amorim_2014}, we found out that the diagrams
from Figure~\ref{fig:Sunset/Fock}(a) and (b) lead to a change of
the bending rigidity $\kappa\sim k^{-\eta}$, and have found that
$\eta=2$ at perturbative level, just like in the classical theory\cite{NP87}.
The difference with respect to the classical case, is that while in
the classical theory a self-consistent calculation (neglecting the
correction to the in-plane elastic constants) changes this value from
$\eta=2$ to $\eta=1$\cite{NP87}, in our zero temperature calculation,
the $\eta=2$ value remained unchanged when doing a similar calculation.

Notice that Kats and Lebedev acknowledge in Ref.~\onlinecite{Kats_2013}
that the diagrams of the form of Figure~\ref{fig:Sunset/Fock}(a)
and (b) lead to $k^{2}$ contributions to the self-energy of the out-of-plane
mode that will be UV (ultraviolet) divergent, as we have found in
Ref.~\onlinecite{Amorim_2014}. Kats and Lebedev argue that such a $k^{2}$
term is a tension and therefore should be forced to be zero, since
a free membrane has zero tension. However, in order to make such a
term zero, one would have to add to the bare action an in-plane tension
term that would act as a counter term. One should notice, that differently from what is done in renormalized perturbation theory, in a Wilsonian renormalization calculation no contour terms are added to the bare action. Furthermore, since the $k^2$ term also appears in the classical theory, we do not see any reason why we should add such an in-plane tension term to the theory and instead take the $k^2$ term as a result of the model, which we interpret to be renormalization of the bending rigidity of the membrane.

\section{Conclusion}

In conclusion, we insist that the $k^{2}$ term in the self-energy of the out-of-plane mode is not a tension term, but a momentum dependent correction to the bending rigidity of the membrane.  A $k^{2}$ term is also found at perturbative level in the classical theory of membranes\cite{NP87}. Therefore, such a term should not be forced to be zero. We should emphasize, however, that the discussion on this note concerns only a perturbative calculation, which is the approach used both in our work Ref.~\onlinecite{Amorim_2014} and by Kats and Lebedev in Ref.~\onlinecite{Kats_2013}. We know that in the classical theory a complete understanding of the physics of membranes requires a non-perturbative treatment of interactions\cite{NPW04}. Such a non-perturbative treatment will also be necessary in the quantum case.

Furthermore, the use of a high momentum cutoff,
and its identification with the Debye momentum, allows us to estimate
the contribution from high momentum modes and is enough to make the
kind of estimations we do in our paper Ref.~\onlinecite{Amorim_2014}. Therefore, we consider the criticisms made by Kats and Lebedev to be unjustified.

\begin{acknowledgments}
B.A. acknowledges support from
Funda\c{c}\~{a}o para a Ci\^{e}ncia e a Tecnologia (Portugal), through Grant.
No. SFRH/BD/78987/2011.  R.R. acknowledges financial support from the Juan de la Cierva Programe
(MEC, Spain).  
E.C. acknowledges support from the European
project FP7-PEOPLE-2013-CIG "LSIE\_2D" and Italian
National MIUR Prin project 20105ZZTSE.
R.R. and F.G. thank financial support from MINECO, Spain, through Grant No. FIS2011-23713. M.I.K and A.F. acknowledge 
funding from the European Union Seventh 
Framework Programme under grant agreement n604391 Graphene Flagship.
\end{acknowledgments}


\begin{thebibliography}{13}%
\makeatletter
\providecommand \@ifxundefined [1]{%
 \@ifx{#1\undefined}
}%
\providecommand \@ifnum [1]{%
 \ifnum #1\expandafter \@firstoftwo
 \else \expandafter \@secondoftwo
 \fi
}%
\providecommand \@ifx [1]{%
 \ifx #1\expandafter \@firstoftwo
 \else \expandafter \@secondoftwo
 \fi
}%
\providecommand \natexlab [1]{#1}%
\providecommand \enquote  [1]{``#1''}%
\providecommand \bibnamefont  [1]{#1}%
\providecommand \bibfnamefont [1]{#1}%
\providecommand \citenamefont [1]{#1}%
\providecommand \href@noop [0]{\@secondoftwo}%
\providecommand \href [0]{\begingroup \@sanitize@url \@href}%
\providecommand \@href[1]{\@@startlink{#1}\@@href}%
\providecommand \@@href[1]{\endgroup#1\@@endlink}%
\providecommand \@sanitize@url [0]{\catcode `\\12\catcode `\$12\catcode
  `\&12\catcode `\#12\catcode `\^12\catcode `\_12\catcode `\%12\relax}%
\providecommand \@@startlink[1]{}%
\providecommand \@@endlink[0]{}%
\providecommand \url  [0]{\begingroup\@sanitize@url \@url }%
\providecommand \@url [1]{\endgroup\@href {#1}{\urlprefix }}%
\providecommand \urlprefix  [0]{URL }%
\providecommand \Eprint [0]{\href }%
\providecommand \doibase [0]{http://dx.doi.org/}%
\providecommand \selectlanguage [0]{\@gobble}%
\providecommand \bibinfo  [0]{\@secondoftwo}%
\providecommand \bibfield  [0]{\@secondoftwo}%
\providecommand \translation [1]{[#1]}%
\providecommand \BibitemOpen [0]{}%
\providecommand \bibitemStop [0]{}%
\providecommand \bibitemNoStop [0]{.\EOS\space}%
\providecommand \EOS [0]{\spacefactor3000\relax}%
\providecommand \BibitemShut  [1]{\csname bibitem#1\endcsname}%
\let\auto@bib@innerbib\@empty
\bibitem [{\citenamefont {Amorim}\ \emph {et~al.}(2014)\citenamefont {Amorim},
  \citenamefont {Rold\'an}, \citenamefont {Cappelluti}, \citenamefont
  {Fasolino}, \citenamefont {Guinea},\ and\ \citenamefont
  {Katsnelson}}]{Amorim_2014}%
  \BibitemOpen
  \bibfield  {author} {\bibinfo {author} {\bibfnamefont {B.}~\bibnamefont
  {Amorim}}, \bibinfo {author} {\bibfnamefont {R.}~\bibnamefont {Rold\'an}},
  \bibinfo {author} {\bibfnamefont {E.}~\bibnamefont {Cappelluti}}, \bibinfo
  {author} {\bibfnamefont {A.}~\bibnamefont {Fasolino}}, \bibinfo {author}
  {\bibfnamefont {F.}~\bibnamefont {Guinea}}, \ and\ \bibinfo {author}
  {\bibfnamefont {M.~I.}\ \bibnamefont {Katsnelson}},\ }\href {\doibase
  10.1103/PhysRevB.89.224307} {\bibfield  {journal} {\bibinfo  {journal} {Phys.
  Rev. B}\ }\textbf {\bibinfo {volume} {89}},\ \bibinfo {pages} {224307}
  (\bibinfo {year} {2014})}\BibitemShut {NoStop}%
\bibitem [{\citenamefont {Nelson}\ and\ \citenamefont {Peliti}(1987)}]{NP87}%
  \BibitemOpen
  \bibfield  {author} {\bibinfo {author} {\bibfnamefont {D.}~\bibnamefont
  {Nelson}}\ and\ \bibinfo {author} {\bibfnamefont {L.}~\bibnamefont
  {Peliti}},\ }\href {\doibase 10.1051/jphys:019870048070108500} {\bibfield
  {journal} {\bibinfo  {journal} {J. Phys. (Paris)}\ }\textbf {\bibinfo
  {volume} {48}},\ \bibinfo {pages} {1085} (\bibinfo {year}
  {1987})}\BibitemShut {NoStop}%
\bibitem [{\citenamefont {Landau}\ and\ \citenamefont {Lifshitz}(1959)}]{LL59}%
  \BibitemOpen
  \bibfield  {author} {\bibinfo {author} {\bibfnamefont {L.}~\bibnamefont
  {Landau}}\ and\ \bibinfo {author} {\bibfnamefont {E.}~\bibnamefont
  {Lifshitz}},\ }\href@noop {} {\emph {\bibinfo {title} {Course of Theoretical
  Physics vol. 7: "Theory of Elasticity"}}}\ (\bibinfo  {publisher} {Pergamon
  Press},\ \bibinfo {address} {Oxford},\ \bibinfo {year} {1959})\BibitemShut
  {NoStop}%
\bibitem [{\citenamefont {{Kats}}\ and\ \citenamefont
  {{Lebedev}}(2014)}]{KL_14c}%
  \BibitemOpen
  \bibfield  {author} {\bibinfo {author} {\bibfnamefont {E.~I.}\ \bibnamefont
  {{Kats}}}\ and\ \bibinfo {author} {\bibfnamefont {V.~V.}\ \bibnamefont
  {{Lebedev}}},\ }\href@noop {} {\bibfield  {journal} {\bibinfo  {journal}
  {ArXiv e-prints}\ } (\bibinfo {year} {2014})},\ \Eprint
  {http://arxiv.org/abs/1407.4298} {arXiv:1407.4298 [cond-mat.stat-mech]}
  \BibitemShut {NoStop}%
\bibitem [{\citenamefont {Kats}\ and\ \citenamefont
  {Lebedev}(2014)}]{Kats_2013}%
  \BibitemOpen
  \bibfield  {author} {\bibinfo {author} {\bibfnamefont {E.~I.}\ \bibnamefont
  {Kats}}\ and\ \bibinfo {author} {\bibfnamefont {V.~V.}\ \bibnamefont
  {Lebedev}},\ }\href {\doibase 10.1103/PhysRevB.89.125433} {\bibfield
  {journal} {\bibinfo  {journal} {Phys. Rev. B}\ }\textbf {\bibinfo {volume}
  {89}},\ \bibinfo {pages} {125433} (\bibinfo {year} {2014})}\BibitemShut
  {NoStop}%
\bibitem [{\citenamefont {Fasolino}\ \emph {et~al.}(2007)\citenamefont
  {Fasolino}, \citenamefont {Los},\ and\ \citenamefont {Katsnelson}}]{FLK07}%
  \BibitemOpen
  \bibfield  {author} {\bibinfo {author} {\bibfnamefont {A.}~\bibnamefont
  {Fasolino}}, \bibinfo {author} {\bibfnamefont {J.}~\bibnamefont {Los}}, \
  and\ \bibinfo {author} {\bibfnamefont {M.~I.}\ \bibnamefont {Katsnelson}},\
  }\href {\doibase 10.1038/nmat2011} {\bibfield  {journal} {\bibinfo  {journal}
  {Nature materials}\ }\textbf {\bibinfo {volume} {6}},\ \bibinfo {pages} {858}
  (\bibinfo {year} {2007})}\BibitemShut {NoStop}%
\bibitem [{\citenamefont {Zakharchenko}\ \emph {et~al.}(2009)\citenamefont
  {Zakharchenko}, \citenamefont {Katsnelson},\ and\ \citenamefont
  {Fasolino}}]{ZKF09}%
  \BibitemOpen
  \bibfield  {author} {\bibinfo {author} {\bibfnamefont {K.~V.}\ \bibnamefont
  {Zakharchenko}}, \bibinfo {author} {\bibfnamefont {M.~I.}\ \bibnamefont
  {Katsnelson}}, \ and\ \bibinfo {author} {\bibfnamefont {A.}~\bibnamefont
  {Fasolino}},\ }\href {\doibase 10.1103/PhysRevLett.102.046808} {\bibfield
  {journal} {\bibinfo  {journal} {Phys. Rev. Lett.}\ }\textbf {\bibinfo
  {volume} {102}},\ \bibinfo {pages} {046808} (\bibinfo {year}
  {2009})}\BibitemShut {NoStop}%
\bibitem [{\citenamefont {Nelson}\ \emph {et~al.}(2004)\citenamefont {Nelson},
  \citenamefont {Piran},\ and\ \citenamefont {Weinberg}}]{NPW04}%
  \BibitemOpen
  \bibinfo {editor} {\bibfnamefont {D.}~\bibnamefont {Nelson}}, \bibinfo
  {editor} {\bibfnamefont {T.}~\bibnamefont {Piran}}, \ and\ \bibinfo {editor}
  {\bibfnamefont {S.}~\bibnamefont {Weinberg}},\ eds.,\ \href@noop {} {\emph
  {\bibinfo {title} {Statistical Mechanics of Membranes and Surfaces}}}\
  (\bibinfo  {publisher} {World Scientific, Singapore},\ \bibinfo {year}
  {2004})\BibitemShut {NoStop}%
\bibitem [{\citenamefont {Xing}\ \emph {et~al.}(2003)\citenamefont {Xing},
  \citenamefont {Mukhopadhyay}, \citenamefont {Lubensky},\ and\ \citenamefont
  {Radzihovsky}}]{XML03}%
  \BibitemOpen
  \bibfield  {author} {\bibinfo {author} {\bibfnamefont {X.}~\bibnamefont
  {Xing}}, \bibinfo {author} {\bibfnamefont {R.}~\bibnamefont {Mukhopadhyay}},
  \bibinfo {author} {\bibfnamefont {T.~C.}\ \bibnamefont {Lubensky}}, \ and\
  \bibinfo {author} {\bibfnamefont {L.}~\bibnamefont {Radzihovsky}},\ }\href
  {\doibase 10.1103/PhysRevE.68.021108} {\bibfield  {journal} {\bibinfo
  {journal} {Phys. Rev. E}\ }\textbf {\bibinfo {volume} {68}},\ \bibinfo
  {pages} {021108} (\bibinfo {year} {2003})}\BibitemShut {NoStop}%
\bibitem [{\citenamefont {Braghin}\ and\ \citenamefont
  {Hasselmann}(2010)}]{BH10}%
  \BibitemOpen
  \bibfield  {author} {\bibinfo {author} {\bibfnamefont {F.~L.}\ \bibnamefont
  {Braghin}}\ and\ \bibinfo {author} {\bibfnamefont {N.}~\bibnamefont
  {Hasselmann}},\ }\href {\doibase 10.1103/PhysRevB.82.035407} {\bibfield
  {journal} {\bibinfo  {journal} {Phys. Rev. B}\ }\textbf {\bibinfo {volume}
  {82}},\ \bibinfo {pages} {035407} (\bibinfo {year} {2010})}\BibitemShut
  {NoStop}%
\bibitem [{\citenamefont {Hasselmann}\ and\ \citenamefont
  {Braghin}(2011)}]{Hasselmann_2011}%
  \BibitemOpen
  \bibfield  {author} {\bibinfo {author} {\bibfnamefont {N.}~\bibnamefont
  {Hasselmann}}\ and\ \bibinfo {author} {\bibfnamefont {F.~L.}\ \bibnamefont
  {Braghin}},\ }\href {\doibase 10.1103/PhysRevE.83.031137} {\bibfield
  {journal} {\bibinfo  {journal} {Phys. Rev. E}\ }\textbf {\bibinfo {volume}
  {83}},\ \bibinfo {pages} {031137} (\bibinfo {year} {2011})}\BibitemShut
  {NoStop}%
\bibitem [{\citenamefont {Cowley}(1963)}]{C__63}%
  \BibitemOpen
  \bibfield  {author} {\bibinfo {author} {\bibfnamefont {R.~A.}\ \bibnamefont
  {Cowley}},\ }\href@noop {} {\bibfield  {journal} {\bibinfo  {journal} {Adv.
  Phys.}\ }\textbf {\bibinfo {volume} {12}},\ \bibinfo {pages} {421} (\bibinfo
  {year} {1963})}\BibitemShut {NoStop}%
\bibitem [{\citenamefont {Katsnelson}\ and\ \citenamefont
  {Trefilov}(2002)}]{KT_02}%
  \BibitemOpen
  \bibfield  {author} {\bibinfo {author} {\bibfnamefont {M.~I.}\ \bibnamefont
  {Katsnelson}}\ and\ \bibinfo {author} {\bibfnamefont {A.~V.}\ \bibnamefont
  {Trefilov}},\ }\href@noop {} {\emph {\bibinfo {title} {Dynamics and
  Thermodynamics of Crystal Lattices}}}\ (\bibinfo  {publisher} {Atomizdat},\
  \bibinfo {address} {Moscow},\ \bibinfo {year} {2002})\BibitemShut {NoStop}%
\end{thebibliography}

%

\end{document}